# Attention-Based Acoustic Feature Fusion Network for Depression Detection

Xiao Xu[1], Yang Wang[2], Xinru Wei[1], Fei Wang[2, 3]*, Xizhe Zhang[1]*

[1]School of Biomedical Engineering and Informatics, Nanjing Medical University, Nanjing, P.R. China.

[2]Early Intervention Unit, Department of Psychiatry, The Affiliated Brain Hospital of Nanjing Medical University, Nanjing, P.R. China.

[3]The Fourth School of Clinical Medicine, Nanjing Medical University, Nanjing, P.R. China.

*: Correspondence: Fei Wang: fei.wang@yale.edu, Xizhe Zhang: zhangxizhe@njmu.edu.cn

**Abstract:**

Depression, a common mental disorder, significantly influences individuals and imposes considerable societal impacts. The complexity and heterogeneity of the disorder necessitate prompt and effective detection, which nonetheless, poses a difficult challenge. This situation highlights an urgent requirement for improved detection methods. Exploiting auditory data through advanced machine learning paradigms presents promising research directions. Yet, existing techniques mainly rely on single-dimensional feature models, potentially neglecting the abundance of information hidden in various speech characteristics. To rectify this, we present the novel Attention-Based Acoustic Feature Fusion Network (ABAFnet) for depression detection. ABAFnet combines four different acoustic features into a comprehensive deep learning model, thereby effectively integrating and blending multi-tiered features. We present a novel weight adjustment module for late fusion that boosts performance by efficaciously synthesizing these features. The effectiveness of our approach is confirmed via extensive validation on two clinical speech databases, CNRAC and CS-NRAC, thereby outperforming previous methods in depression detection and subtype classification. Further in-depth analysis confirms the key role of each feature and highlights the importance of MFCC-related features in speech-based depression detection.

# 1. Introduction

Depression, a widespread mental disorder, has severe negative impacts on both physical and mental health, and can even lead to loss of life. With the outbreak of COVID-19, depression cases have risen, putting a significant financial burden on society [1]. At present, diagnosing depression largely relies on clinical assessments, which are time-consuming, subjective, and heavily dependent on the physicians' expertise [2]. Main diagnostic tools used to measure severity, like the Hamilton Depression Rating Scale (HAMD) [3], depending on interviews carried out by doctors or self-reports by patients, which scores the patient's behavior. The growing number of depression cases have put a lot of pressure on diagnostic accuracy; therefore, an objective, machine learning-based method could greatly improve the detection of Major Depressive Disorder (MDD).

Diagnosing depression comes with major challenges, mainly due to the lack of diagnostic methods and the inherent subjectivity in depression detection [4]. To overcome these challenges, there is a growing interest in using behavioral markers for automated depression diagnosis and stage prediction [5], [6]. By recording and tracking these behavioral markers, we can potentially develop both screening tools and treatment options for those suffering from depression. Recently, numerous automatic depression estimation (ADE) methodologies have emerged, using machine learning and audio-visual techniques to measure depression severity based on audio-visual cues [7]–[10]. Such methods aim to support clinicians in diagnosing depression severity using audio-visual information [11]. Furthermore, they offer many benefits over traditional diagnostic methods, including being non-invasive, objective, and scalable, making them very appealing for both clinical practice and research applications.

The idea of using voice data for depression detection is supported by advancements in artificial intelligence. These methods can be looked at from various perspectives. For instance, many understandable speech features can be calculated by classic signal processing techniques and used in thorough analyses correlating with depression [12]. Within this feature category, studies involving voice quality [13] to spectrum correlation have shown great promise [14]. Some deep learning-based key works, such as DepAudioNet [15] et al., provide a complete representation of speech signal characteristics. This is achieved by building deep learning models using Long Short-Term Memory (LSTM) and Convolutional Neural Networks (CNN) among others.

Depression detection through speech analysis is a complex task that requires the careful consideration of various speech features and modeling algorithms. Despite the wealth of research in this area, several challenges persist. Firstly, the potential of deep learning-based speech feature fusion methods remains largely untapped [17]. Numerous features have been proven to extract information carried by original speech. By fusing these features, we can extract more comprehensive emotional information from speech. Secondly, the efficacy of multi-feature fusion techniques, particularly in terms of efficient feature extraction and fusion strategies, warrants further exploration. Compared to single-feature methods, fusion strategies that combine different features have demonstrated significant performance improvements in detecting depression [17], [18]. Thirdly, the features extracted from speech are diverse in representation and size, necessitating tailored modeling for each feature. Therefore, it is imperative to develop a reliable depression detection model that leverages speech feature fusion. Lastly, given that speech is a time-series signal, the interactions of long time-series signals should be considered in the processing of speech features.

The above challenges lead us to two key questions: (1) Which speech features are most effective for detecting depression, and how can we extract them to create pre-trained models that adequately capture crucial information in time-series signals without loss during computation? (2) How can we execute feature fusion to demonstrate a clear improvement over the use of a single feature, and what strategies should we employ to model these diverse features effectively?

To tackle the above challenges, we propose a novel acoustic feature fusion approach: the Attention-Based Acoustic Feature Fusion Network (ABAFnet) to detect depression through speech. In addressing the first question, we first train our technical strategy to incorporate deep feature vectors for the effective extraction and processing of complex information inherent in speech signals. We choose a variety of commonly used speech features as the inputs to our neural network. These speech features are then transformed into deep feature vectors, which serve as the final features in our pre-trained models. To tackle the second question, we present a late fusion strategy, which aids us in combining and optimizing multiple features to improve depression detection. Specifically, we evaluate the performance of each individual feature within its sub-model and adjust each feature's weight dynamically before combining them into a final 1-dimensional feature. This fusion approach not only optimizes the combined features but also facilitates the effective modeling of these diverse features. Our study identified specific speech features that can indicate depression symptoms, emphasizing the innovativeness of Speech-based Depression Detection (SDD) tasks with ABAFnet. Overall, our study highlights the potential of automatic depression detection systems as invaluable tools in supporting mental health care and improving the quality of life for individuals affected by depression.

In summary, the main contributions of our paper are as follows:

1) We propose the Attention-Based Acoustic Feature Fusion Network (ABAFnet), a deep learning model that utilizes a speech feature fusion strategy for depression detection. Our approach focuses on achieving information complementarity among features, enabling the extraction of additional depression-related acoustic markers, which significantly enhances the effectiveness of depression detection. In addition, we employ a late fusion approach to integrate the pre-trained features. Furthermore, we incorporate an Attention Mechanism and LSTM to capture the time-series information inherent in the features, further strengthening the model's ability to understand temporal dynamics and contextual information. This comprehensive approach improves the accuracy and reliability of depression detection.

2) We have compiled a new Chinese Clinical Neutral Reading Audio Corpus (CNRAC) with the reading text, "Let Life be Beautiful like Summer Flowers," and a new Chinese Neutral Reading Audio Corpus from a mental health screening among College Students (CS-NRAC) with the reading text, "The North Wind and the Sun." These databases served as the basis for training and evaluating our model.

The rest of the paper is organized as follows: Section 2 discusses related works. In Section 3, our proposed method and network architecture are presented in detail. Next, experiments and the results conducted on two databases are presented in Section 4. Finally, some discussions and future works are given.

## 2. Related Work

In the following content, we explore the existing real-world research and the significance of diagnosing depression through voice in a clinical setting. Additionally, we delve into the evolution of various voice features, the development of voice-based deep learning models, and the progress in fusion technologies.

Acoustic biomarkers refer to vocal characteristics and patterns that can be indicative of an individual's mental health state. These can range from pitch, tone, and volume to more complex features like speech rate, variability, and prosody. Several studies [19], [20] have found that individuals with depression often exhibit distinct vocal characteristics. For instance, they might have a more monotone voice, reduced speech variability, or altered speech tempo. These vocal changes are believed to reflect the underlying emotional and cognitive disturbances associated with depression. For example, Mundt et.al [21] examined the speech patterns as potential biomarkers for depression severity and response to treatment. They underscored that vocal acoustic properties could signify depression severity and treatment efficacy, with treatment responders exhibiting distinct vocal changes. Silva et.al [22] assessed voice acoustic parameters in distinguishing between depressed and non-depressed individuals. Emphasizing F0 (SD), jitter, shimmer, and CPPS, the research found these parameters notably varied between groups, with jitter and CPPS acting as depression predictors. The collective findings underscore the potential of vocal acoustic metrics in depression diagnosis and assessment.

Vocal features have shown significant value in research related to depression detection. For instance, the amplitude of sound waves and spectrograms not only reveals the foundational structure of speech but also suggest emotional fluctuations tied to depression [19]. The Mel-spectrogram stands out because it emulates human auditory perception, becoming vital in identifying subtle vocal shifts potentially linked to depression. Similarly, the Constant Q-transform (CQT) and Variable Q-transform (VQT) [23] have demonstrated their worth in depression-associated voice analysis. These tools, through their frequency domain representation, offer researchers a deeper avenue to capture vocal characteristics potentially associated with depressive states. Additionally, with the continuous advancement of technology, one-dimensional features like the i-vector [24], d-vector, and x-vector [25] have been utilized in pinpointing vocal differences connected to depression.

The above-mentioned features are widely used as inputs to SDD tasks, while the deep features are also popular. Firstly, several studies have used CNNs to extract deep features from speech data. For example, in [26], the authors used a refined neural network architecture, specifically the Squeeze-and-Excitation Residual Network (SE-ResNet), to extract detailed spectrogram features. Similarly, in [27], researchers suggested a frame rate-based data augmentation strategy that uses CNNs for deep speech feature extraction. Secondly, a significant number of research focus on LSTM, highlighting the effective use of x-vectors for anxiety and depression detection. Notably, Kwon et al. [28] introduced an innovative approach using x-vectors in a binary classification model for phone recordings, a method confirmed by another study [25] that used x-vectors to extract important speaker characteristics, demonstrating their potential for distinguishing depression. Thirdly, many studies have attempted to combine CNN and LSTM or use alternative techniques. For example, a study by [29] developed a method to analyze both language and non-verbal elements of speech by converting spoken words into feature vectors and using a Bidirectional LSTM (BiLSTM) for comprehensive language and non-verbal modeling. Another study [30] fused multi-modal data using Gated Recurrent Unit (GRU) and BiLSTM to create a multi-modal fusion network from an Emotional Audio-Textual Depression Corpus. Furthermore, several innovative attention-related methodologies have been introduced. A study [31] focused on vowel-level spectrotemporal information using various data augmentation techniques, while the Hierarchical Attention Transfer

Network (HATN) [32] aimed to apply attention mechanisms from speech recognition to assist in measuring depression severity. Lastly, a novel methodology for SDD was proposed by [33], introducing an attention mechanism called Multi-Local Attention (MLA) that was used in conjunction with LSTM, showing excellent performance in scenarios where data was limited.

The field of attention-based feature fusion has seen significant advancements. For instance, in [34], the PGA-Net, a framework for surface defect detection, used a pyramid feature fusion and global context attention network to improve accuracy and prediction. Building on this, [35] introduced the FFA-Net, incorporated a new Feature Attention module, a basic block structure with Local Residual Learning, and an Attention-based Feature Fusion structure. Finally, [36] proposed a comprehensive plan called attentional feature fusion to improve feature fusion, integrating multiscale channel attention modules and iterative attentional feature fusion to address integration bottlenecks, with fewer layers and parameters.

Building upon these studies, our work introduces a distinct approach to depression detection, emphasizing deep feature fusion based on voice data. Unlike past methodologies that may have addressed features and modeling in isolation, we converge deep learning intricacies into a coherent model that effectively harnesses the latent power of vocal nuances. Our work, thus, stands as a testament to the union of rigorous research and innovation, poised to revolutionize depression diagnosis through voice.

## 3. Methods

### 3.1 Problem Definition

In our study, we use the variable $x_i$ to represent the raw speech data of the $i$-th subject, where $x_i$ is a real number in $R^{L_{(i)}}$. $L_{(i)}$ represents the length of the speech data, and I represents the total number of subjects. Our objective is to detect potential signs of depression by extracting features from the speech data and assigning a label $y_i$ to each subject. The label $y_i$ belongs to the set $\{0, 1\}$ and indicates the respective depression state.

The process of our SDD task consists of the following stages:

1. **Preprocessing**: The raw speech $x_i$ undergoes a preprocessing sequence that includes Voice Activity Detection (VAD) and downsampling. The outcome of this stage is denoted as $x_{i,p}$ and is a real number in $R^{L'_{(i)}}$, where $L'_{(i)}$ is less than or equal to $L_{(i)}$. $x_{i,p}$ serves as a concise representation of the original speech.

2. **Feature Extraction**: From $x_{i,p}$, two types of primary features are extracted: image-based features ($F_{img,i}$) and numerical vector features ($F_{num,i}$). This transformation is described by a function $G_\alpha: R^{L'_{(i)}} \to R^{K \times J}$, where $K$ and $J$ define the dimensions of the image and vector features respectively. $\alpha$ represents the learnable parameters.

3. **Weight Adjustment**: The Weight Adjustment Module (WAM) denoted as $\mathcal{W}_\beta(\cdot)$ is used to combine the extracted features. β represents the parameters set that are optimized during this process. The combined feature representation is denoted as $f''_i$, which is then passed through an LSTM-Attention pipeline, resulting in a transformed feature $f'''_i$.

The overall goal is to jointly optimize the parameters $\alpha$, $\beta$, and the parameters of the classifier $\gamma$ to solve the classification challenge. Formally, we aim to find the optimal values for $\alpha$, $\beta$, and $\gamma$ by minimizing the loss function $\mathcal{L}\left(C_\gamma\left(W_\beta\left(G_\alpha(x_{(i,p)})\right)\right), y_i\right)$ for all $(x, y)$ pairs in the dataset $D$. Here, $C_\gamma$ represents the classifier and $\mathcal{L}$ represents the loss function.

Our primary ambition is to deduce the optimal values for $\alpha$, $\beta$, and $\gamma$ through iterative refinement.

### 3.2 Speech preprocess and feature extraction

In this subsection, we will describe our preprocessing pipeline and feature extraction process, which is shown in **Figure 1**, our preprocessing process begins with recordings that adhere to a uniform standard, both in terms of recording equipment and script. Using VAD, we extract the pure vocal segments, eliminating all extraneous sounds. Following this, we employ downsampling techniques to reduce the computational complexity in subsequent steps. In the feature extraction phase, we extract four types of features, including the upper envelope, spectrogram, Mel-spectrogram, and HSFs.

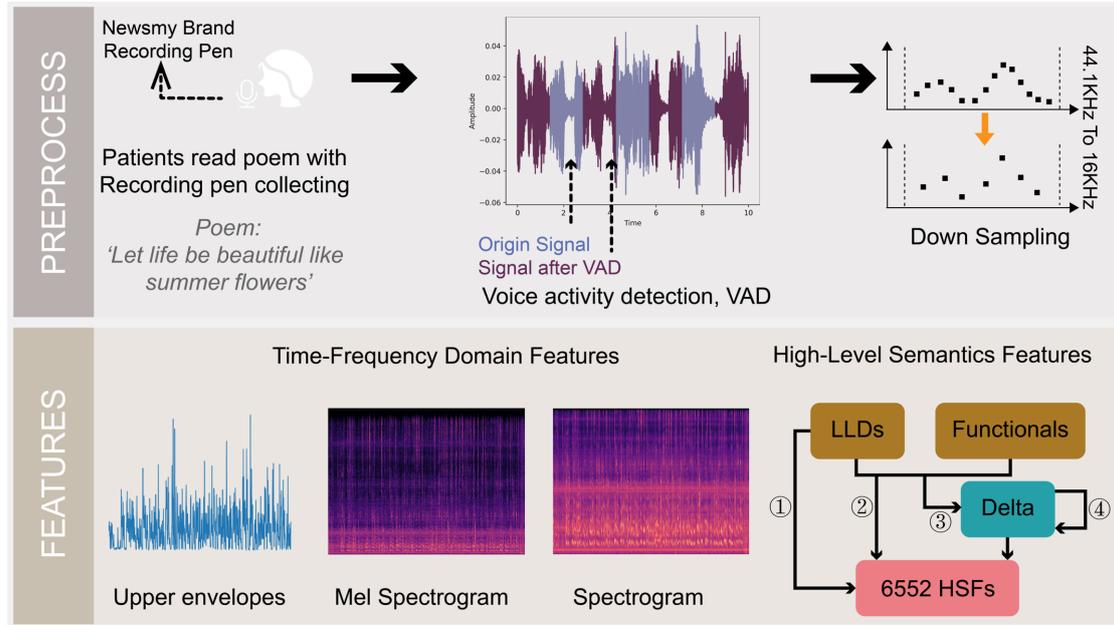

**Figure 1. Pre-processing and feature extraction.**

#### 3.2.1 Audio Preprocessing

Our audio preprocessing pipeline is split into two main stages: VAD and Resampling.

**1. VAD:** Our study focuses on the vocal component, requiring the removal of non-vocal parts of the recordings to reduce interference with subsequent investigations. To this end, we use a dual-threshold endpoint detection algorithm to cut and retain relevant sections. This algorithm operates by evaluating the short-term energy and zero-crossing rate of the audio signal. The implementation sequence of the algorithm is outlined as follows:

Initially, we denote $s(n)$ as the input signal, where $n$ represents the time index. We then calculate the short-time energy (STE) using a Hamming window, denoted as $w(n)$. The STE can be determined as:

$$E(n) = \sum_{m=-\infty}^{+\infty} s^2(m) \cdot w(n-m)$$

The start and end of speech segments can be approximated as:

$$n_{start} = \min_{n}(n|E(n) > HT), n_{end} = \min_{n>n_{start}}(n|E(n) < LT)$$

where $HT$ and $LT$ are the high and low thresholds, respectively.

**2. Resampling:** The original clinical recordings had a sampling rate of 44.1 kHz. To lessen computational demands in the following analyses and remove high-frequency noise, we used the FFMPEG tool, known for audio and video processing, to downsample the audio to 16 kHz. This rate is frequently used in audio signal analysis and effectively maintains the quality of speech signals [37].

### 3.2.2 Feature Extraction

We utilize the Librosa [38] and OpenSMILE [39] libraries to extract features.

**1. Upper Envelope:** This feature of a signal outlines the smooth curve mapping its peaks. We used a simple approach to compute the envelope of a signal $x(n)$ by performing an absolute value operation followed by a low-pass filter:

$$E(n) = abs(x(n)) \cdot h(n)$$

Where $n$ is the time index, $h(n)$ is the impulse response of a low-pass filter.

**2. Spectrogram:** The spectrogram was calculated by dividing the signal into overlapping frames, windowing each frame (using a Hanning window of 2048 samples in this case), and conducting the Fast Fourier Transform (FFT) on each frame. The formula for the FFT is:

$$X(k) = \sum_{n=0}^{N-1} x(n) * e^{-j\frac{2\pi kn}{N}}$$

Where $x(n)$ is the time-domain signal, $N$ is the total number of samples, $j$ is the imaginary unit, $k$ is the frequency index (denoting the specific frequency being considered), and $X(k)$ is the frequency-domain signal, calculated by applying the FFT to $x(n)$.

**3. Mel-spectrogram:** We first compute the magnitude spectrum of the signal using FFT. Next, we map the powers of the spectrum obtained in the first step onto the Mel scale using overlapping triangular windows. Finally, we take the logarithms of the powers at each Mel frequency. The relation between Mel frequency and actual frequency is given by:

$$m = k * \log_{10}(1 + \frac{f}{700})$$

Where $k$ is a scaling constant that makes the Mel scale align with the frequency in Hz at lower frequencies, $f$ is the frequency in Hz, $m$ is the perceived frequency in Mel, and the number 700 approximates the human ear's response to different frequencies.

**4. HSFs:** Using OpenSMILE, an audio feature extraction tool, we computed the emoLarge speech feature set, an emotion-related HSF speech feature [38]. This has been widely used in studies focusing on emotion recognition and affective computing. A total of 6552 features, corresponding to their respective Low-Level Descriptors (LLDs) and Functionals (as summarized in **Table 1**), were extracted from the emoLarge speech feature set in this study.

**Table 1. Overview of the emoLarge dataset.**

| | Group |
|---|---|

| 2 Energy Related LLDs | |
|---|---|
| LOGenergy | Prosodic |
| zero-crossing rate | Prosodic |
| **50 Spectral Related LLDs** | |
| fband 0-250, 0-650, 230-650, 1000-4000 | Spectral |
| spectralRollOff 25.0, 50.0, 75.0, 90.0 | Spectral |
| spectral flux, centroid, maxPos, minPos | Spectral |
| mfcc[0-12] | Cepstral |
| melspec[0-25] | Spectral |
| **3 Voicing Related LLDs** | |
| voiceProb | Sound quality |
| F0, F0env | Prosodic |
| **Functionals Applied To LLD/LLD-delta/LLD-delta-delta** | |
| maxPos, minPos | Locations |
| numPeaks, meanPeakDist, peakMean, peakMeanMeanDist | Peaks |
| range, amean, absmean, qmean, nzabsmean, nzqmean, nzgmean, nnz | Basic Statistics |
| quartile1, quartile2, quartile3, iqr1-2, iqr2-3, iqr1-3 | Quartile Statistics |
| percentile95.0, percentile98.0 | Percentile Statistics |
| centroid, variance, stddev, skewness, kurtosis | Central Moment |
| zcr | Rate |
| linregc1, linregc2, linregerrA, linregerrQ | Distance Regression |
| qregc1, qregc2, qregc3, qregerrA, qregerrQ | Secondary Regression |
| maxameandist, minameandist | Prosodic |

### 3.3 Attention-Based Acoustic Feature Fusion Network

Our study aims to boost model prediction capabilities by integrating multiple speech features. This could potentially yield more comprehensive information about speech patterns associated with depression. In this section, we initially present an overview of the framework, followed by an in-depth discussion of ABAFnet.

### 3.3.1 Overview of Framework

Our study commences with the preprocessing of raw speech data, symbolized as $x_i$, from the i-th subject. This speech is in the space $R^{L_{(i)}}$, where $L_{(i)}$ indicates the length of the $i$-th subject's speech $x_i \in R^{L_{(i)}}$. After preprocessing the raw data, we extract four types of features: three image-based features and one numerical statistical feature. These features are subsequently amalgamated, denoted as $f_{img,i}, f_{num,i} \in R^{L_{(i)}}, f_i = [f_{img,i}, f_{num,i}]$.

The architecture of our framework is shown in **Figure 2**. We utilize a dual-phase fusion method powered by the WAM. Initially, features are directed into two distinct sub-models: the ImageModel, which manages image features, and the NumModel, which processes the numerical features. The results from these models influence the WAM, paving the way for the subsequent fusion step. In the second stage, the features, now adjusted by the output of the first stage, are amalgamated in the WAM, thus performing a late fusion. Subsequent to this fusion, the weighted feature vector $f_i''$ is input into the LSTM-Attention channel, resulting in the final feature representation $f_i''' = \text{LSTM} - \text{Att}(f_i'')$.

In conclusion, we jointly optimize the parameters derived from all stages, along with the classifier parameters, to confront the classification challenge. We aim to minimize the loss $\mathcal{L}$ between the actual and predicted depression states (notated as $y_i$ and $\hat{y}_i$, respectively) over the dataset $\min_{\theta,c} E_{(x,y)\in D}\mathcal{L}\left(C_c\left(f_i'''\right), y_i\right)$. Using the WAM, our feature fusion strategy aims to enhance both the extraction of relevant information and the performance of the classifier, thereby efficiently identifying depression based on speech features.

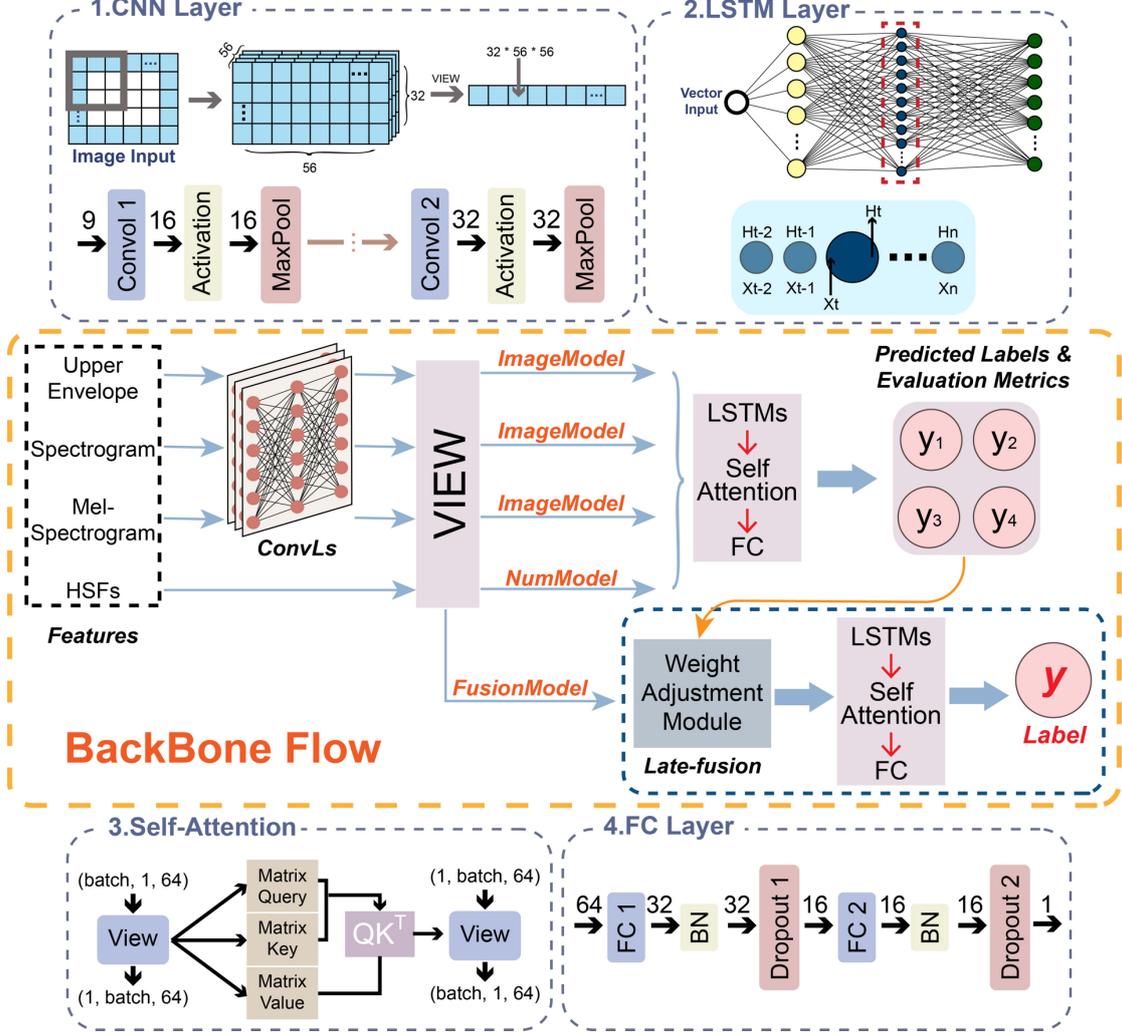

**Figure 2**. Detailed architecture of ABAFnet.

### 3.3.2 Features Pre-training

To ensure the effectiveness of each combined feature, we use two customized modules, referred to as ImageModel and NumModel, to handle three image-centric features and one 1-dimensional HSFs, respectively. Each feature goes through individual training and testing within these sub-models to collect evaluation metrics. The features that contribute to effective classification are then retained for integration into the composite model.

In ImageModel, we use CNNs to extract structured hierarchical features from images, which require minimal preprocessing. Drawing from the methodologies detailed in [40] and [41], we first use CNNs to extract depth features from image-category speech features. These features are then expanded into one-dimensional vectors to be combined with other speech features. The equation that encapsulates this CNN module is:

$$Out_{conv}^{N \times F_1 \times H_1 \times W_1} = \sum_{m=-1}^{1} \sum_{n=-1}^{1} \sum_{c=0}^{2} I^{N \times C \times H \times W}[i+m, j+n, c] * K[m+1, n+1, c, k]$$

Where $N, F_1, H_1, W_1$ denote the batch size, number of filters, and the spatial dimensions of the output feature maps respectively. $I^{N \times C \times H \times W}$ is the input image tensor of shape ($N$, $C$, $H$, $W$), with $C$ representing the number of channels, and $H$, $W$ standing for the image's height and width respectively. $K$ represents the kernel tensor of the convolution operation. $N, F_1, H, W, C \in Z^+$, $m, n \in \{-1, 0, 1\}, i \in \{0, \ldots, H-1\}, j \in \{0, \ldots, W-1\}, c \in \{0, \ldots, C-1\}, k \in \{0, \ldots, F_1-1\}$.

The resulting one-dimensional tensor is then processed through the LSTM model. As demonstrated in [42] and [43] during the LSTM's process of reducing dimensionality, the model can extract higher-level features from speech data. To improve our model's handling of temporal information, we incorporated Attention Mechanisms, which enhance performance by understanding context and dependencies [44], [45] [46]. Specifically, [47] and [48] illustrate the potential of sub-attention in language modeling and enhancement. These collective findings underscore the powerful potential of Attention Mechanisms in speech signal processing. The LSTM-Attention module is described by the following equation:

Let us define:
$$O = Out_{conv}^{N \times T \times D}$$
$$L(O, t) = \text{LSTM}(O[t-1], O[t])$$

Then, our equation becomes:
$$Out_{lstm+attn}^{N \times T \times D} = \sum_{t=1}^{T} \alpha_t L(O, t)$$

$$\text{Where } \alpha_t = \frac{\exp(L(O,t))}{\sum_{t=1}^{T} \exp(L(O,t))}$$

Where $Out_{lstm+attn}^{N \times T \times D}$ denotes the output from the LSTM-Attention layer, a tensor of shape ($N$, $T$, $D$) where $N$ is the batch size, $T$ is the sequence length, and $D$ is the hidden size of the LSTM layer. $D \in Z^+$, $t \in \{-1, \ldots, T\}$.

Subsequently, the output from the Attention Mechanism is processed through a series of Fully Connected (FC) layers, with dropout and batch normalization operations interspersed for optimal performance. The dropout layers function to mitigate the risk of overfitting [49], [50]. This FC module is defined by the following equation:

Let's define:
$$X_1 = \sigma_1(W_1 * Out_{lstm+attn}^{N \times T \times D} + b_1)$$
$$X_2 = \sigma_2(W_2 * \text{dropout}(X_1) + b_2)$$

Then, our final FC module output is:
$$Out_{fc} = \sigma_3(W_3 * \text{dropout}(X_2) + b_3)$$

Where $W_1, b_1, W_2, b_2, W_3, b_3$ are weights and biases of FC, $\sigma_i$ are the activation functions.

In the case of the NumModel, it can be seen as an ImageModel without the CNN module. However, the key difference can be summarized as follows:

$$Out_{fc}^{N \times D_{out}} = FC\left(\text{LSTM} + \text{Attn}\left(I_{model}^{N \times T \times D_{input}}\right)\right)$$

Here, the key difference lies in $I_{model}$. For the ImageModel, $I_{model} = Out_{pool2}^{N \times 32 \times 56 \times 56}.\text{reshape}(N, 1, 32 * 56 * 56)$, representing the reshaped output from the second pooling layer. For the NumModel, $I_{model} = I^{N \times H \times W \times C}.\text{reshape}(N, 1, D_{input})$, which is the reshaped original input.

### 3.3.3 Late Fusion Strategy

In the field of research focused on multiple modes of data, the chosen strategy for merging these data types plays a crucial role, as it substantially affects the overall performance of the model. Both the work [18], which combines data during the information gathering phase and a model that merges all data at the initial stage [51], have used early fusion methodologies and thus outperformed single-mode approaches. However, considering the unique characteristics of our dataset, late fusion emerges as a more appropriate choice for our experimental design. Many comparable studies have initially trained separate modes of data in deep learning models before merging the results from various model outputs [52], [53]. The main advantage of late fusion lies in its ability to combine the scores of each mode at the decision-making level, thus allowing the model to consider the unique contribution of each mode and allocate weights based on its relevance to the task at hand. In addition, late fusion can correct alignment differences between modes, which is crucial given the asynchronous nature of our multiple features. Once the fusion strategy is finalized, we plan to incorporate WAM to optimize the weights of the four input features and build the highest-performing model.

In the WAM, we have introduced several evaluation metrics, depending on the objective of the task. The WAM is calculated as follows:

$$Score_i = \alpha * Acc_i + \beta * Precision_i + \gamma * Recall_i + \delta * MacroAvg_i + \epsilon * WeightedAvg_i$$

The weight for sub-model i ($w_i$) is calculated as the ratio of its score ($Score_i$) to the sum of the scores of all sub-models ($\sum_j Score_j$):

$$w_i = \frac{Score_i}{\sum_j Score_j}$$

Here, $\alpha, \beta, \gamma, \delta, \epsilon$ are the weight parameters to adjust according to the task requirements. If overall performance is important, we can set $\alpha$ higher than the other weights, for instance, $\alpha = 0.5$, and the rest all being 0.1. If recognizing positive samples matters more, we can set $\beta$ and $\gamma$ higher than the other weights, like $\beta = \gamma = 0.3$, and the rest all being 0.1. If model robustness is the concern, we can set $\delta$ and $\epsilon$ higher than the other weights, for instance, $\delta = \epsilon = 0.3$, and the rest all being 0.1.

This method allows us to dynamically assign importance to each sub-model based on its performance, supporting the effective integration of diverse input features within the ABAFnet.

### 3.3.4 Training parameters

When the parameter count in a model notably exceeds the volume of training data, there's a risk that the model may excessively learn from the training dataset, capturing noise and anomalies—a prevalent issue. Our strategy encompasses designing a model with minimal essential layers to constrain the parameter count while also adopting Dropout and Early Stopping techniques to mitigate overfitting risks. The parameter count for our model design is detailed in **Table 2**.

**Table 2. Parameters overview.**

| Layer | Discription | Total parameters |
|---|---|---|
| CNN | CNN with two Conv2D layers (16 and 32 filters) and one FC layer (128 neurons) | ~12.85M |
| FC | FC layer (128 neurons) taking a 291-dimension numeric vector | 37,376 |
| Attention | Attention mechanism with an embed dimension of 256 and 4 heads | 1,048,576 |

## 4. Experiments

This section presents a set of experiments conducted on two datasets to assess the effectiveness of ABAFnet. The experiments aim to answer the following research questions:

1. How does ABAFnet compare in performance to other current top-performing models?

2. Are the features we have chosen significantly relevant?

3. What are the primary acoustic features associated with depression?

**4.1 Datasets**

**1. CNRAC:** This corpus was curated at the Early Intervention Department of The Affiliated Brain Hospital of Nanjing Medical University. The dataset systematically investigates the differences in speech variations between individuals diagnosed with depression and normal controls (NCs). It comprises data from 155 patients with depression and 216 individuals classified as NCs, based on the Hamilton Depression Rating Scale (HAMD-17) [3]. The HAMD-17, a clinician-administered instrument, measures the severity of depression on a scale ranging from 0-7 (normal) to > 23 (very severe).

Initially, patients were recruited from the inpatient and outpatient units of the Affiliated Brain Hospital of Nanjing Medical University. NCs without a history of depression were recruited through online advertisements. The inclusion criteria encompassed regardless of sex, normal hearing, and proficiency in Mandarin. The exclusion criteria included organic mental disorders or other psychiatric conditions, medically diagnosed diseases that may affect vocalization and substance-induced mental disorders.

The group of patients with depression consists of 39 males and 116 females, aged 13-24 years (M = 15, SD = 2). In contrast, the NC group includes 78 males and 138 females, aged 10-48 years (M = 27, SD = 6). Clinical data were compiled following the DSM-IV (Diagnostic and Statistical Manual of Mental Disorders, Fourth Edition) guidelines [54].

During the data collection phase, all participants were positioned in a quiet and enclosed environment. Clear instructions were provided to the participants, emphasizing the importance of maintaining a natural and relaxed state throughout the recording session. Participants were asked to read the specific text "Let Life be Beautiful like summer flowers." The collected audio data are stored in WAV format with a sampling rate of 44.1 kHz, and PCM at 16-bits. The speech lengths range from 2 to 4 minutes. The data were captured using a standardized Newsmy brand recording pen.

**2. CS-NRAC:** This larger corpus was collated at Nanjing Medical University during September and October 2022. This collection involved 1561 medical freshmen with an average age of 18 years, who were all healthy individuals, partaking in psychological screening.

Like the previous methodology, participants were asked to complete the Patient Health Questionnaire-9 (PHQ-9) [55] via the WeChat official account platform. The PHQ-9 is a self-administered tool for diagnosing depression, where score interpretation ranges from 1-4 (minimal) to 20-27 (severe). It is important to note that while PHQ-9 aids in the identification of depressive symptoms, it does not result in a clinical diagnosis. In this phase, participants read the fable "The North Wind and the Sun," within a controlled quiet environment. The audio data was initially recorded in M4A format with an 8 KHz sampling frequency. Each recording, approximately 60 seconds long, was subsequently converted to WAV format files sampled at 16 KHz to maintain consistency with **CNRAC**.

We applied a stringent screening process for the university student dataset. This was driven by the understanding that data could be unreliable if the average response time was less than 1.5 seconds or more than 20 seconds, or if the total reading time did not reach 40 seconds. Endpoint detection was utilized to identify human voices and maintain the quality of the data. Recordings that did not adhere to certain criteria, such as those with excessive background noise or voice distortion, were regarded as dirty data and were consequently excluded. This exclusion criterion was enforced to ensure that only clean and relevant data contributed to the analysis, thus ensuring the robustness of the dataset.

**Table 3** and **Table 4** show the details of datasets.

**Table 3. Detail of CNRAC on HAMD-17.**

| CNRAC | NC | Depression | | |
| --- | --- | --- | --- | --- |
| | | Mild | Moderate | Severe |
| Score | 0-7 | 8-16 | 17-24 | ≥25 |
| Number | 216 | 13 | 62 | 80 |

**Table 4. Detail of CS-NRAC on PHQ-9.**

| CS-NRAC | NC | Depression | | | | |
| --- | --- | --- | --- | --- | --- | --- |
| | | Minimal | Mild | Moderate | Moderately Severe | Severe |
| Score | 0 | 1-4 | 5-9 | 10-14 | 15-19 | ≥20 |
| Number | 786 | 565 | 190 | 16 | 4 | 0 |

## 4.2 Comparison of Single and Fusion Features

To assess the efficacy of the ABAFnet in the SDD task, we initiated a thorough experiment on the CNRAC dataset using the HAMD-17 scale. Our evaluation began by validating the effectiveness of individual features. We examined four distinct features independently within our ImageModel and NumModel. Subsequently, these features were integrated into the FusionModel to evaluate the overall performance of our model. Our evaluation methodology incorporated a five-fold cross-validation and employed an early stopping strategy with a patience threshold of 10. All experimental procedures were executed on a Tesla A100 GPU utilizing PyTorch.

In **Table 5** and **Figure 3**, we present a comparative analysis of the performance of our fusion strategy against single-feature predictions. For a consistent comparison across model modules, every sub-model, inclusive of the FusionModel, was constructed utilizing our developed sub-modules. The results showed that our fusion approach demonstrated superior performance with an ACC of 0.814 and AUC of 0.847, which outperforms other single-feature approaches. This suggests our method's proficiency in amalgamating diverse feature information, showcasing its enhanced performance in SDD tasks.

**Table 5. Evaluation metrics of each model.**

| Models | ACC | Precision | Recall | F1 | ROC-AUC |
| --- | --- | --- | --- | --- | --- |
| Upper Envelope | 0.642 ± 0.054 | 0.516 ± 0.101 | 0.567 ± 0.142 | 0.521 ± 0.064 | 0.658 ± 0.059 |
| Spectrogram | 0.685 ± 0.063 | 0.608 ± 0.091 | 0.727 ± 0.096 | 0.654 ± 0.060 | 0.762 ± 0.063 |
| Mel-spectrogram | 0.752 ± 0.049 | 0.657 ± 0.072 | 0.769 ± 0.095 | 0.701 ± 0.032 | 0.798 ± 0.032 |
| HSFs | 0.641 ± 0.018 | 0.461 ± 0.032 | 0.426 ± 0.056 | 0.463 ± 0.018 | 0.559 ± 0.046 |
| **Fusion** | **0.814 ± 0.041** | **0.650 ± 0.090** | **0.800 ± 0.133** | **0.700 ± 0.030** | **0.847 ± 0.044** |

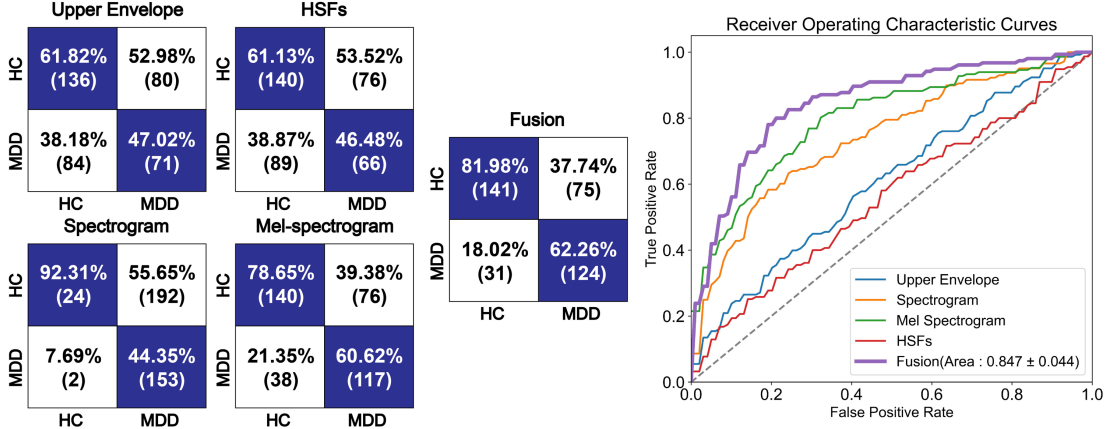

**Figure 3. Confusion matrix and ROC curve.** The result plots include the confusion matrix results for the four sub-model and the confusion matrix, along with their ROC plots.

### 4.3 Performance Evaluation on CNRAC

Following preliminary binary tests centered on detecting the presence or absence of depression, we broadened the scope of our investigation to encompass varying degrees of depression severity. We stratified depression severity based on the HAMD-17 scales, segmenting it into four clearly defined levels: NC (Normal Control), Mild, Moderate, and Severe depression, as shown in Table 4. Due to the imbalanced distribution of subjects across these categories, random downsampling was employed to equilibrate the dataset. This experiment was iterated 10 times, and the metrics from each iteration were averaged to ensure robustness in our results.

Our experimental results, as detailed in **Table 6**, show the strong performance of our models in classifying depression severity. In the NC-Subtype classifications, where we distinguished between NC and various levels of severity, our model achieved high accuracy rates ranging from 93.4% to 97.5%. While precision scores exhibited variability, the consistently elevated recall rates (ranging from 84.7% to 86.0%) underscore the model's prowess in correctly identifying true positive instances, thereby emphasizing its adeptness in depression detection. Notably, classifications involving NC and Moderate severity showed exceptional performance across all measured metrics.

For the Inter-Subtype classifications, which differentiated between various levels of depression severity, accuracy rates ranged from 85.5% and 93.5%. While these are slightly lower compared to NC-Subtype, these results still affirm the model's effectiveness. This claim is further supported by the excellent performance on precision, recall, F1 scores, and AUC values, which together emphasize our model's capability in distinguishing different degrees of depression severity.

**Table 6. Results of Subtype Classification.**

| Tasks | Measures | ACC | Precision | Recall | F1 | AUC |
|---|---|---|---|---|---|---|
| NC-Subtype | NC vs Mild | 0.975 ± 0.021 | 0.251 ± 0.267 | 0.860 ± 0.237 | 0.314 ± 0.244 | 0.864 ± 0.175 |
| | NC vs Moderate | 0.934 ± 0.033 | 0.697 ± 0.181 | 0.847 ± 0.186 | 0.749 ± 0.157 | 0.946 ± 0.041 |
| | **NC vs Severe** | **0.935 ± 0.032** | **0.716 ± 0.220** | **0.853 ± 0.190** | **0.731 ± 0.162** | **0.940 ± 0.082** |
| Inter-Subtype | Mild vs Moderate | 0.855 ± 0.190 | 0.762 ± 0.350 | 0.637 ± 0.388 | 0.662 ± 0.361 | 0.790 ± 0.217 |
| | **Mild vs Severe** | **0.935 ± 0.050** | **0.846 ± 0.271** | **0.654 ± 0.355** | **0.701 ± 0.322** | **0.814 ± 0.213** |
| | Moderate vs Severe | 0.862 ± 0.064 | 0.826 ± 0.083 | 0.815 ± 0.162 | 0.806 ± 0.110 | 0.902 ± 0.068 |

### 4.4 Ablation Experiment

Carrying out an ablation study is a crucial step when a model is built on multiple features, as shown in our study where we used four separate features. This analysis allows us to understand the individual impact of each feature on the overall performance of the model, providing valuable insights into how the model works. The ultimate aim is to determine the importance of each feature, which can guide future efforts to optimize feature selection and improve the effectiveness of the model. For this experiment, we adjust WAM in which each feature is sequentially left out while the remaining three are averaged and inputted into the fusion model.

Our ablation study, as depicted in **Table 7** and **Figure 4**, reveals the importance of each feature in contributing to the effectiveness of our model. When a single feature is excluded, the accuracy rates ranged from 60.9% (when Mel-spectrogram was excluded) to 76.9% (when HSFs were excluded). The corresponding Precision, Recall, F1, and AUC metrics also showed variability. Notably, the removal of HSFs resulted in a decrease in the model's performance, with an accuracy of 76.9% and an AUC of 0.775, which are significantly lower than the full model's performance. This finding highlights the indispensable role of HSFs, emphasizing the unique contribution of each feature to the robustness and overall performance of the model. These insights will guide future efforts to further optimize feature selection, underscoring the need to incorporate a diverse set of features to achieve the best predictive efficiency.

**Table 7. Results of Ablation Experiment.**

| Exclude | ACC | Precision | Recall | F1 | AUC |
| --- | --- | --- | --- | --- | --- |
| Upper Envelope | 0.620 ± 0.021 | 0.350 ± 0.186 | 0.613 ± 0.448 | 0.398 ± 0.249 | 0.548 ± 0.040 |
| Spectrogram | 0.615 ± 0.011 | 0.405 ± 0.035 | 0.671 ± 0.304 | 0.475 ± 0.144 | 0.499 ± 0.059 |
| Mel-spectrogram | 0.609 ± 0.020 | 0.243 ± 0.207 | 0.400 ± 0.400 | 0.282 ± 0.252 | 0.501 ± 0.100 |
| HSFs | 0.769 ± 0.062 | 0.721 ± 0.223 | 0.600 ± 0.321 | 0.549 ± 0.084 | 0.775 ± 0.149 |

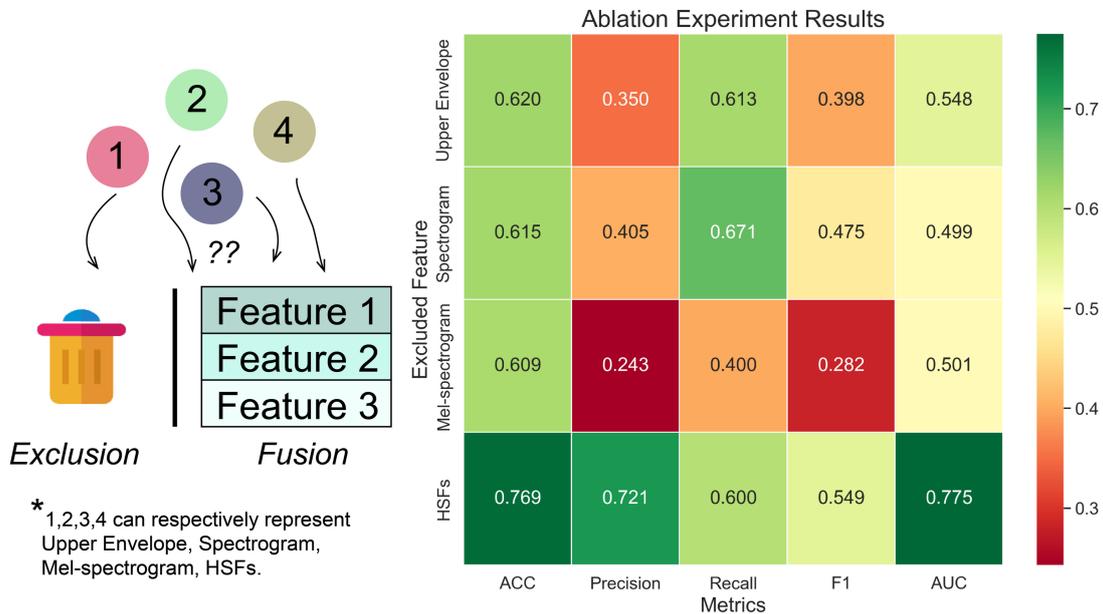

**Figure 4. Ablation experiment flow and heatmap results.** Red means high importance and green means low importance.

### 4.5 Analysis of Interpretable HSFs

Our previous experiment has already confirmed the usefulness of HSFs in binary depression prediction, as shown by the performance of our HSFs sub-model. As part of our ongoing efforts to understand intricate patterns of depressive states from speech signals and identify potential acoustic markers, we conducted more detailed analyses.

To investigate the HSFs, we used two-sample T-tests with False Discovery Rate (FDR) correction, a method commonly used to compare the means of two independent sample groups. We used a threshold of p-value of 0.01 to select the 32 features with significant differences. At the same time, for binary classification, we used a Random Forest algorithm, which helped us identify significant features and their corresponding importance scores.

Our findings, as outlined in **Table 8**, show notable differences especially in the outcomes of Mel Frequency Cepstral Coefficients (MFCC) and Mel Spectrum related features. Out of the 32 distinct features explored, the MFCC category constitutes a significant 43.75% (14 features), the same as the Mel Spec. Collectively, these two pivotal categories represent a substantial 87.5% of the total features examined, a testament to their importance in the SDD task. The standout features, mfcc_sma_de_de[0]_quartile2 and pcm_fftMag_melspec_sma[1]_meanPeakDist, accentuates the influence of the two category, separately. Our findings not only illuminate the key role these features play in identifying acoustic markers for depression but also solidify their consistency with results from our previous experiments.

**Table 8. Significant Differences HSFs.**

| FeatureName | F | FDR | RF_Score | Category |
|---|---|---|---|---|
| pcm_fftMag_melspec_sma[1]_meanPeakDist | -3.31 | 9.44E-03 | 1.68E-02 | Mel Spec |
| mfcc_sma_de_de[0]_quartile2 | -5.58 | 1.32E-05 | 1.44E-02 | MFCC |
| pcm_fftMag_melspec_sma_de_de[0]_meanPeakDist | -4.75 | 1.35E-04 | 1.39E-02 | Mel Spec |
| pcm_zcr_sma_de_de_minPos | -4.21 | 7.53E-04 | 1.02E-02 | Envelope |
| mfcc_sma[7]_numPeaks | -5.18 | 5.11E-05 | 8.14E-03 | MFCC |
| pcm_fftMag_melspec_sma_de[0]_meanPeakDist | -4 | 1.41E-03 | 7.94E-03 | Mel Spec |
| pcm_fftMag_melspec_sma_de_de[15]_meanPeakDist | -3.3 | 9.44E-03 | 7.13E-03 | Mel Spec |
| pcm_fftMag_melspec_sma_de[16]_meanPeakDist | -4.3 | 6.07E-04 | 7.12E-03 | Mel Spec |
| pcm_fftMag_melspec_sma[1]_minPos | -4.3 | 6.07E-04 | 6.71E-03 | Mel Spec |
| mfcc_sma[7]_skewness | -3.5 | 6.03E-03 | 6.66E-03 | MFCC |
| mfcc_sma[8]_skewness | -3.88 | 1.94E-03 | 6.57E-03 | MFCC |
| pcm_fftMag_melspec_sma[25]_minPos | -4.05 | 1.22E-03 | 6.23E-03 | Mel Spec |
| pcm_fftMag_melspec_sma_de[10]_meanPeakDist | -3.36 | 8.66E-03 | 6.13E-03 | Mel Spec |
| mfcc_sma[5]_numPeaks | -4.92 | 1.01E-04 | 6.05E-03 | MFCC |
| pcm_fftMag_melspec_sma_de[17]_meanPeakDist | -4.68 | 1.61E-04 | 5.61E-03 | Mel Spec |
| mfcc_sma[10]_numPeaks | -4.77 | 1.35E-04 | 5.30E-03 | MFCC |
| pcm_zcr_sma_de_de_skewness | -3.29 | 9.44E-03 | 5.22E-03 | Envelope |
| pcm_fftMag_melspec_sma_de[23]_meanPeakDist | -3.54 | 5.53E-03 | 5.14E-03 | Mel Spec |
| pcm_fftMag_melspec_sma_de[14]_meanPeakDist | -3.71 | 3.34E-03 | 5.13E-03 | Mel Spec |
| pcm_fftMag_spectralRollOff90.0_sma_minPos | -3.42 | 7.17E-03 | 5.07E-03 | Energy Spec |
| mfcc_sma[4]_meanPeakDist | -3.96 | 1.44E-03 | 5.02E-03 | MFCC |
| mfcc_sma[2]_linregc1 | 3.46 | 6.45E-03 | 4.98E-03 | MFCC |
| pcm_fftMag_melspec_sma[2]_minPos | -3.79 | 2.59E-03 | 4.87E-03 | Mel Spec |
| mfcc_sma[2]_numPeaks | -4.9 | 1.01E-04 | 4.85E-03 | MFCC |
| mfcc_sma_de[0]_numPeaks | -4.22 | 7.53E-04 | 4.52E-03 | MFCC |
| mfcc_sma[5]_minPos | -3.97 | 1.44E-03 | 4.46E-03 | MFCC |
| pcm_fftMag_melspec_sma_de_de[23]_meanPeakDist | -3.3 | 9.44E-03 | 4.22E-03 | Mel Spec |
| pcm_fftMag_spectralRollOff75.0_sma_minPos | -4.07 | 1.22E-03 | 3.78E-03 | Energy Spec |
| mfcc_sma[4]_numPeaks | -3.69 | 3.39E-03 | 3.77E-03 | MFCC |
| pcm_fftMag_melspec_sma_de_de[3]_meanPeakDist | -3.68 | 3.41E-03 | 3.53E-03 | Mel Spec |
| mfcc_sma[6]_numPeaks | -4.62 | 1.84E-04 | 3.50E-03 | MFCC |
| mfcc_sma[10]_minPos | -3.49 | 6.03E-03 | 2.97E-03 | MFCC |

## 4.6 Model Comparison
### 4.6.1 Model Performance

To verify the robustness of our proposed ABAFnet model, we set out to contrast it with several deep learning models relevant to the task of detecting depression through speech signals. Since our model blends Convolutional Neural Networks (CNN) and Long Short-Term Memory networks (LSTM), we selected comparable models based on either CNN, LSTM, or a combination of both. These models span from fundamental to advanced techniques. We used the torchaudio library for uniform preprocessing of the raw speech, which served as input for these models. Here's a brief overview of the models considered:

- *Graves et al.* [56] deployed Bidirectional LSTM (**BiLSTM**) for classification tasks, which showed proficiency in managing sequential data and modeling complex temporal relationships.
- *Graves et al.* proposed **StackedLSTM** [57], which uses deep recurrent neural networks (RNNs) with multiple layers of LSTM cells to boost acoustic modeling performance in speech recognition.
- *Sak et al.* introduced **DeepLSTM** [58], a large-scale LSTM-based RNN architecture intended to improve acoustic modeling. This model significantly enhanced speech recognition accuracy.
- *Ma et al.* presented **DepAudioNet** [15], an effective deep model for audio-based depression classification. It employs a mix of CNNs and RNNs to process and analyze audio signals.
- *Wei et al.* proposed **ConvBiLSTM** [59], a model that extracts both short and long-term temporal and spectral features through a hierarchical CNN and BiLSTM structure, enabling end-to-end depression estimation.

Comparing these models, as is shown in **Table 9**, our proposed ABAFnet demonstrated superior performance. Among models based on deep learning methods, ABAFnet achieved notable improvements. For instance, compared to the CNN-based model DepAudioNet, our method showed a 16% increase in F1-score. Similarly, when contrasted with LSTM-based models like BiLSTM and StackedLSTM, ABAFnet outperformed them, attaining an increase in accuracy by 17% and 26%, respectively.

Furthermore, ABAFnet shone even when compared to models integrating both CNNs and LSTMs, such as ConvBiLSTM, with our model achieving a 19% improvement in F1-score. These substantial gains underscore the effectiveness of the ABAFnet framework and its superiority in detecting depression using speech features. Importantly, these advancements are largely attributed to ABAFnet's comprehensive feature extraction capacity, which meticulously dissects and utilizes relevant spectral features associated with depression in high-dimensional signals.

**Table 9. Classification metrics of each prediction model.**

| Model Name | ACC | Precision | Recall | F1-Score |
|---|---|---|---|---|
| BiLSTM | 0.64 | 0.46 | 0.44 | 0.45 |
| StackedLSTM | 0.55 | 0.47 | 0.50 | 0.48 |
| DeepLSTM | 0.57 | 0.71 | 0.26 | 0.38 |
| DepAudioNet | 0.65 | 0.80 | 0.42 | 0.55 |
| ConvBiLSTM | 0.78 | 0.41 | 0.91 | 0.57 |
| **ABAFnet** | **0.81** | **0.65** | **0.80** | **0.70** |

### 4.6.2 Complexity analysis

While our primary focus was on feature-based processing work, it was necessary to consider the time-efficiency in real-world applications compared to end-to-end deep learning models. As such, we tested the running time of the various models on the same hardware.

As represented in **Figure 5**, ABAFnet required a longer training duration. This additional time can be attributed to the model's need to process four distinct features extracted from raw audio signals, instead of using the preprocessed raw audio signals by torchaudio, which was the case for the other models. This extra time spent reading and converting these multi-features subsequently increased the model's complexity and extended the training period.

In terms of model complexity, models like BiLSTM, StackedLSTM, and DeepLSTM, which are built solely on LSTM layers, and DepAudioNet, which relies only on CNN layers, exhibited the shortest training periods. On the other hand, ConvBiLSTM, which integrates both CNN and LSTM layers, showed a noticeable increase in its training duration.

Previous work [60] has shown that feature extraction from raw audio signals tends to be more time-consuming. However, while designing ABAFnet, our focus remained on keeping the parameters to a minimum while ensuring the extraction of detailed and differentiating acoustic features. Despite the increased training time being a result of the added complexity, it's considered reasonable given the overall superior performance of our model's evaluation metrics.

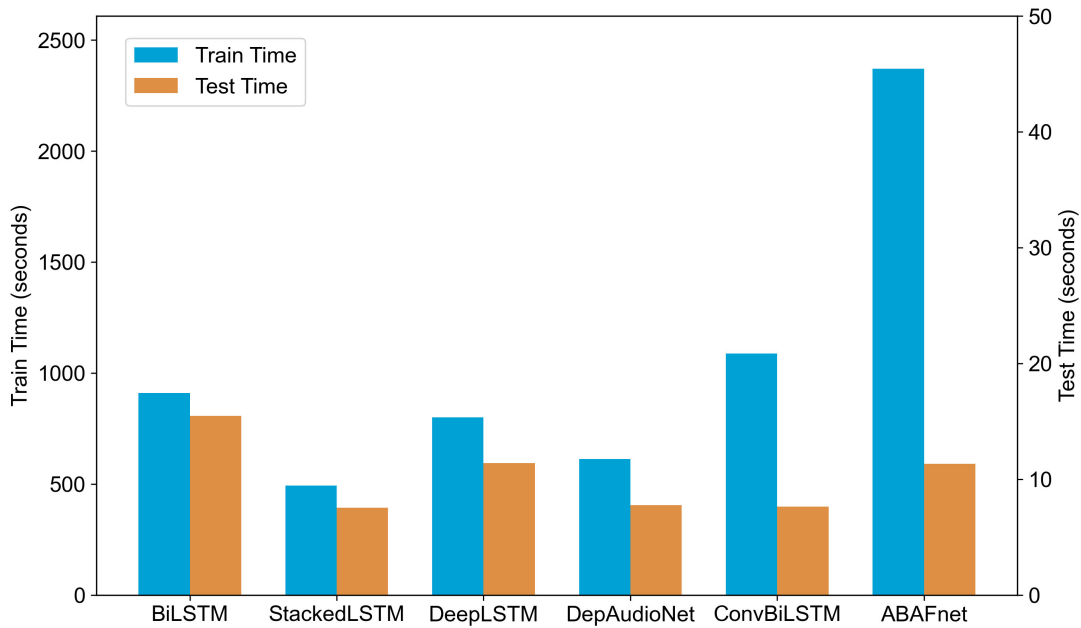

**Figure 5. Running time of the models on CNRAC.** Train time and test time are on the left and right y-axis respectively.

### 4.7 Validate on CS-NRAC

We also employed our ABAFnet model to explore its potential for early detection of depression in large-scale screenings.

We categorized the CS-NRAC scale data based on the PHQ-9 scores. However, considering several influential studies [61], it became clear that the threshold division of PHQ-9 may have certain limitations in terms of confidence and reliability. We chose the left and right endpoint threshold of 4 and 8 in the "Mild" category as our research target. We conducted classification prediction experiments using ABAFnet for threshold values ranging from 3 to 5 and from 8 to 10, we define these two ranges separately as 'Mild Left' and 'Mild Right'. This approach aimed at increasing the accuracy of our model while ensuring its reliable performance across different PHQ-9 score ranges. The dataset was imbalanced, with fewer samples as the scores increased – a common challenge in machine learning applications.

As illustrated in **Table 7** and **Figure 6**, for the 'Mild Left', our model displayed optimal best performance at thresholds 3 and 5, as reflected by all metrics, there is a slight improvement compared to threshold 4. High recall rates of 0.852 and 0.857 suggest the model's impressive sensitivity in identifying the mild depression group within these thresholds, which is particularly significant considering the lower prevalence of higher-scoring samples. On the other hand, for the 'Mild Right', while the model exhibited the highest accuracy at a threshold of 10, it showed zero recall, precision, and F1 score, indicating an inability to accurately classify any true positive instances at this level. This could be partially attributed to the scarcity of higher score samples, which might make the model less effective at detecting the less common but crucial instances of depression. The PR_AUC values support these findings, showing superior performance at lower thresholds within the 'Mild Left' category.

**Table 7. Results of validation on CS-NRAC.**

| Endpoints | Threshold | ACC | Precision | Recall | F1 | PR_AUC |
|---|---|---|---|---|---|---|
| Mild Left | 3 | 0.565 | 0.556 | 0.852 | 0.673 | 0.546 |
|  | 4 | 0.545 | 0.537 | 0.765 | 0.631 | 0.475 |
|  | 5 | 0.587 | 0.568 | 0.857 | 0.684 | 0.541 |
| Mild Right | 8 | 0.537 | 1.000 | 0.050 | 0.095 | 0.471 |
|  | 9 | 0.542 | 0.529 | 0.750 | 0.621 | 0.374 |
|  | 10 | 0.600 | 0.000 | 0.000 | 0.000 | 0.572 |

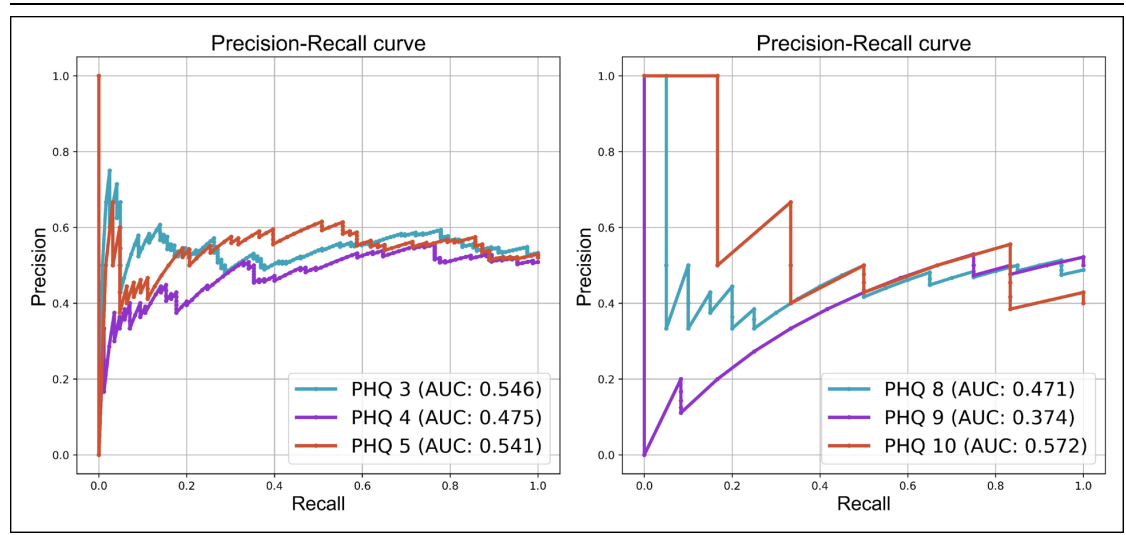

**Figure 6. Model performance on CS-NRAC.** The PR_AUC is used to evaluate the performance of models on datasets where one class of samples is underrepresented.

# 5. Conclusion

In this study, we have introduced ABAFnet, an innovative deep learning framework tailored for depression detection in speech. ABAFnet distinguishes itself through the integration of an attention-driven multi-feature fusion mechanism. Our findings underscore the capability of the framework, with its custom-designed CNN layer adeptly transforming speech signals into tensor representations. The harmonious integration of the LSTM and Attention layers has further cemented ABAFnet's superiority in processing sequential data. This integration, in synergy with the Fusion module embedded with the WAM, dynamically refines the prominence of the features, bolstering the model's performance. Experimental results reveal ABAFnet's robustness, with it consistently outperforming contemporary state-of-the-art models in both the detection of depression and in the nuanced classification of its subtypes. Such results not only attest to the architectural prowess of ABAFnet but also emphasize the potential it holds in the realm of clinical applications.

Looking forward, we are poised to further investigate the intricacies of ABAFnet, with a keen interest in understanding its foundational theory. Moreover, the prospect of leveraging ABAFnet in clinical settings is particularly promising. As the landscape of deep learning continues to evolve, we believe that the advancements within the sub-modules of our model will pave the way for even more optimized and innovative solutions in the future.

## Declarations

**Role of the funding source** This study was funded by the National Natural Science Foundation of China (62176129 to Xizhe Zhang), National Science Fund for Distinguished Young Scholars (81725005 to Fei Wang), NSFC-Guangdong Joint Fund (U20A6005), Jiangsu Provincial Key Research and Development Program (BE2021617 to Fei Wang).
**Data availability statement** Data used in this study is available by contacting the corresponding author to make arrangements. The CNRAC and the CS-NRAC obtained informed consent from all participants, whose ethical clearances were from the Ethics Committee of The Affiliated Brain Hospital of Nanjing Medical University (2021-KY108-01) and the Nanjing Medical University Ethics Committee ((2022)793), respectively. The code repository address is https://github.com/xuxiaoooo/ABAFnet.
**Acknowledgments** The authors would like to acknowledge the subjects, researchers, and instructors for their dedication and help.